\documentclass{elsart}

\usepackage{amssymb,graphicx}

\begin{document}

\begin{frontmatter}


\title{Center of mass energy and system-size dependence of photon production at forward rapidity at RHIC}


 \author[uic]{B. I.~Abelev},
 \author[pu]{M. M.~Aggarwal},
 \author[vecc]{Z.~Ahammed},
 \author[jinr]{A.~V.~Alakhverdyants},
 \author[kent]{B. D.~Anderson},
 \author[bnl]{D.~Arkhipkin},
 \author[jinr]{G. S.~Averichev},
 \author[jammu]{S. K.~Badyal},
 \author[mit]{J.~Balewski},
 \author[uic]{O.~Barannikova},
% ...

(STAR Collaboration)


% ...
\address[bnl]{Brookhaven National Laboratory, Upton, New York 11973}
% ...
\address[uic]{University of Illinois at Chicago, Chicago, Illinois 60607}
% ...
\address[jammu]{University of Jammu, Jammu 180001, India}
\address[jinr]{Joint Institute for Nuclear Research, Dubna, 141 980, Russia}
\address[kent]{Kent State University, Kent, Ohio 44242}
% ...
\address[mit]{Massachusetts Institute of Technology, Cambridge, MA 02139-4307}
% ...
\address[pu]{Panjab University, Chandigarh 160014, India}
% ...
\address[vecc]{Variable Energy Cyclotron Centre, Kolkata 700064, India}
% ...


\begin{abstract}
We present the multiplicity and pseudorapidity distributions of photons produced 
in Au+Au and Cu+Cu collisions at $\sqrt{s_{\mathrm {NN}}}$ = 62.4 and 200 GeV. The 
photons are measured in the region $-3.7 < \eta < -2.3$ using the 
photon multiplicity detector in the STAR experiment at RHIC. The number of photons produced 
per average number of participating nucleon pairs increases
with the beam energy and is independent of the collision centrality.
For collisions with similar 
average numbers of participating nucleons the photon multiplicities are observed to be 
similar for Au+Au and Cu+Cu collisions at a given beam energy. The ratios of the number of charged
particles to photons in the 
measured pseudorapidity range are found to be 
1.4 $\pm$ 0.1 and 1.2 $\pm$ 0.1 for 
$\sqrt{s_{\mathrm {NN}}}$ = 62.4 GeV
and 200 GeV, respectively. 
The energy dependence of this ratio could reflect  
varying contributions from baryons to charged particles, while mesons are  
the dominant contributors to photon production in the given  
kinematic region.
The photon pseudorapidity distributions normalized by average number of participating 
nucleon pairs, when plotted as a 
function of $\eta-y_{\mathrm beam}$, are found to follow a longitudinal scaling independent of 
centrality and colliding ion species at both beam energies.
\end{abstract}

\end{frontmatter}

\section{Introduction}

For high energy heavy-ion collisions,
measurements of particle multiplicity provide information on particle production 
mechanisms~\cite{rhicwhitepapers}. Event-by-event fluctuations in the multiplicity of 
produced particles within a thermodynamic picture could be 
related to matter compressibility~\cite{fluc_gen}. The event-by-event correlation 
between photon and charged particle multiplicities can be used to test the predictions 
of formation of disoriented chiral condensates~\cite{dcc}. The
variation of particle density in pseudorapidity ($\eta$) with collision centrality can shed light on
the relative contribution of soft and hard (perturbative QCD jets) processes in particle 
production~\cite{phenixscaling}.
Multiplicity measurements can provide tests of ideas on initial conditions in 
heavy-ion collisions based on parton saturation~\cite{partonsaturation} and 
color glass condensates~\cite{cgc}. Under certain model assumptions, the particle density in 
pseudorapidity can provide information on the initial temperature and velocity of sound in 
the medium~\cite{initial}. The pseudorapidity distributions are found to be sensitive to 
the effects of re-scattering, hadronic final-state interactions, and longitudinal flow~\cite{width}. 

Several interesting features of the dependence of particle density in pseudorapidity have 
been observed in Au+Au collisions from the experiments at the Relativistic Heavy-Ion Collider (RHIC). 
Particle production is found to follow a 
unique, collision energy independent, longitudinal
scaling~\cite{limiting_frag} in $p$+$p$ and $d$+Au, as well 
as in heavy-ion collisions~\cite{phobosscaling,brahms}.
Such longitudinal scaling is also found to be independent of collision centrality 
for photons~\cite{starphoton,pmdftpc}. The total charged particle multiplicity (integrated
over the full pseudorapidity range) per average number of participating nucleon ($\langle N_{\mathrm{part}} \rangle$) 
pair is found to be independent of collision centrality~\cite{phobosscaling}. However, at 
mid-rapidity ($|\eta|$ $<$ 1),
charged particle multiplicity per $\langle N_{\mathrm{part}} \rangle$ is observed to increase from peripheral to central 
collisions~\cite{phobosscaling}. This 
clearly indicates that the mechanism of particle production could be different in different 
pseudorapidity 
regions. 
In the year 2005, a unique opportunity to investigate the system-size dependence of global 
observables occurred when Cu+Cu collisions 
were produced at RHIC.
In light of the earlier results of photon multiplicity 
scaling with $\langle N_{\mathrm{part}} \rangle$~\cite{starphoton,pmdftpc} at forward rapidity, one could
make direct comparison of the observables ($N_{\gamma}$ and $dN_{\gamma}/d\eta$) 
for Cu+Cu and Au+Au collisions for systems having similar values of $\langle N_{\mathrm{part}} \rangle$.

In this paper we present the first measurements of
photon multiplicity distributions
at forward rapidities in Cu+Cu collisions at $\sqrt{s_{\rm{NN}}}$ = 62.4 and 200 GeV and
Au+Au collisions at  $\sqrt{s_{\rm{NN}}}$ = 200 GeV from the STAR experiment~\cite{star_nim} at RHIC. 
The results from Au+Au collisions at $\sqrt{s_{\rm{NN}}}$ = 62.4 GeV were reported in 
Refs.~\cite{starphoton,pmdftpc}.
The photon multiplicity measurements are presented for various
collision centrality classes 
and are compared to corresponding results for charged particles. 
The photon production 
is dominated by those from the decay of $\pi^{0}$s~\cite{starphoton}. 
{\sc hijing}~\cite{hijing} calculations indicate that about 93--96\% of photons 
are from inclusive $\pi^0$ decays for the $\sqrt{s_{\rm{NN}}}$ and $\eta$ range studied. 

\section{Experiment and Analysis}

The STAR detector contains several subsystems which measure hadronic and electromagnetic
observables at forward rapidity~\cite{star_nim}. The main subsystem used in the present analysis is the 
Photon Multiplicity Detector (PMD)~\cite{starpmd_nim}.  
Photons are detected using a highly granular 
preshower PMD located $-5.4$ m from the center of the Time Projection Chamber (TPC), the nominal collision 
point, along the beam axis. The measurements are carried out within the pseudorapidity region 
of $-3.7$ to $-2.3$  at $\sqrt{s_{\rm{NN}}}$ = 62.4 and 200 GeV for Au+Au and Cu+Cu collisions. A minimum
bias trigger is obtained using the charged particle hits from the Central Trigger Barrel (CTB), an array 
of scintillator slats arranged
in a barrel around the TPC, two Zero Degree Hadronic Calorimeters (ZDCs) located $\pm$18 m 
from the center of the TPC, and two Beam-Beam Counters (BBCs)~\cite{trigger}. 
A total of 307k, 334k, 289k and 330k minimum bias events for 
Au+Au 200, Au+Au 62.4, Cu+Cu 200 and Cu+Cu 62.4 GeV collisions, respectively, were analyzed. These events have a 
collision vertex position less than 30 cm 
from the center of the TPC along the beam axis. The centrality determination in this analysis
uses the uncorrected multiplicity of charged particles in the region $|\eta|$ $<$ 0.5, as
measured in the TPC. The average number of participating nucleons 
is obtained from Monte Carlo Glauber calculations~\cite{star_glauber}. The $\langle N_{\mathrm{part}} \rangle$
values
corresponding to various percentages of the cross section for Au+Au and Cu+Cu collisions at
$\sqrt{s_{\rm{NN}}}$ = 62.4 and 200 GeV are given in the Table~\ref{table0}.

\begin{table}
\caption{\label{table0}
Average number of participating nucleons ($\langle N_{\mathrm{part}} \rangle$) for various collision 
centralities for Au+Au and Cu+Cu collisions at $\sqrt{s_{\rm{NN}}}$ = 62.4 and 200 GeV. }
\vspace{.5cm}
\begin{center}
\begin{tabular}{|c|c|c|c|c|c|c|}
\hline
\% cross section & $\langle N_{\rm{part}}^{\rm{AuAu}} \rangle$ & $\langle N_{\rm{part}}^{\rm{AuAu}} \rangle$ & $\langle N_{\rm{part}}^{\rm{CuCu}} \rangle$ & $\langle N_{\rm{part}}^{\rm{CuCu}} \rangle$ \\
      & 200 GeV           & 62.4 GeV          & 200 GeV          &   62.4 GeV           \\
\hline
0-5    & $352.4^{+3.4}_{-4.0}$     & $347.3^{+4.3}_{-3.7}$   &      --               &  --   \\
0-10   & $325.9^{+5.5}_{-4.3}$  &          --             & $99.0^{+1.5}_{-1.2}$  &  $96.4^{+1.1}_{-2.6}$ \\
5-10   & $299.3^{+6.6}_{-6.7}$     & $293.3^{+7.3}_{-5.6}$   &   ---                 &   --  \\
10-20  & $234.5^{+9.1}_{-7.8}$  & $229.0^{+9.2}_{-7.7}$   & $74.6^{+1.3}_{-1.0}$ &  $72.2^{+0.6}_{-1.9}$  \\
20-30  & $166.6^{+10.1}_{-9.6}$ & $162.0^{+10.0}_{-9.5}$  & $53.7^{+1.0}_{-0.7}$ &  $51.8^{+0.5}_{-1.2}$ \\
30-40  & $115.5^{+9.6}_{-9.6}$  & $112.0^{+9.6}_{-9.1}$   & $37.8^{+0.7}_{-0.5}$ &  $36.2^{+0.4}_{-0.8}$ \\
40-50  & $76.7^{+9.0}_{-9.1}$   & $74.2^{+9.0}_{-8.5}$    & $26.2^{+0.5}_{-0.4}$ &  $24.9^{+0.4}_{-0.6}$ \\
50-60  & $47.3^{+7.6}_{-8.1}$   & $45.8^{+7.0}_{-7.1}$    & $17.2^{+0.4}_{-0.2}$ &  $16.3^{+0.4}_{-0.3}$ \\
60-70  & $26.9^{+5.5}_{-6.5}$   & $25.9^{+5.6}_{-5.6}$    &  --                   &  --                    \\
70-80  & $14.1^{+3.6}_{-4.0}$   & $13.0^{+3.4}_{-4.6}$    &  --                   &  --                    \\
\hline
\end{tabular}
\end{center}
\end{table}

The PMD consists of two planes (charged particle veto and preshower)
of arrays of cellular gas proportional counters.
A lead plate (3 radiation lengths thickness) is placed between
the two planes and serves as a photon converter.
The sensitive medium is a gas mixture of Ar and
CO$_2$ in the ratio 
70:30 by weight.
There are 41,472 hexagonal cells in each plane, which are located inside 
12 high voltage insulated and gas-tight chambers called supermodules. 
To each supermodule is applied a negative voltage of 1400 V as the operating
voltage.
A photon traversing the converter produces an electromagnetic shower 
in the preshower plane, leading to a large signal, spread over several 
cells. In contrast, a charged particle's signal is essentially confined 
to a single cell. 
The photon conversion efficiency studied from simulations is found to
increase with increasing photon energy ($E_{\gamma}$) up to 1 GeV and then
saturate for higher energies. The typical values for the three radiation length
converter are observed to be around 70\% for $E_{\gamma}$ = 0.2 GeV
and 95\% for $E_{\gamma}$ = 1 GeV.
In the present analysis, only the data 
from the preshower plane are used. 
Further details of the design
and characteristics of the PMD are found in Ref.~\cite{starpmd_nim}. 

The analysis of the data from the PMD involves the following:
(a) event selection, (b) cell-to-cell gain calibration, and (c) 
reconstruction or extraction of the photon multiplicity. 
The cell-to-cell gain calibration is done using the ADC distributions 
of isolated cells (cells with six neighbouring cells having zero ADC). 
The ADC distribution of an isolated cell is treated 
as the response of the cell to charged particles, corresponding to a minimum ionizing 
particle (MIP)~\cite{starpmd_nim}.
For most of the cells this response follows a Landau distribution. 
We use the mean of the ADC distribution of isolated cells
to estimate and correct the relative gains of all cells within each
supermodule. The cell-to-cell gain variation is between 10--25\%
for different supermodules.
The extraction of the photon multiplicity proceeds in two steps involving  
clustering of hits and photon-hadron discrimination. Hit clusters consist of 
contiguous cell signals. Photons are separated from charged particles using the
following conditions: (a) the number of cells in a cluster is  $>$ 1, and (b) 
the cluster signal is larger than 3 times the average MIP response 
of all isolated cells in a supermodule. The choice of the conditions 
is based on results of detailed simulations~\cite{starphoton,pmdftpc,starpmd_nim}. 
The number of selected clusters, called
$\gamma{\mathrm -like}$  clusters ($N_{\mathrm {\gamma{\mathrm -like}}}$), in different
supermodules for the same $\eta$ coverage are used to evaluate 
the effect of possible
non-uniformity in the response of the detector. 


\begin{figure*}
\begin{center}
\includegraphics[height=10cm,width=12cm]{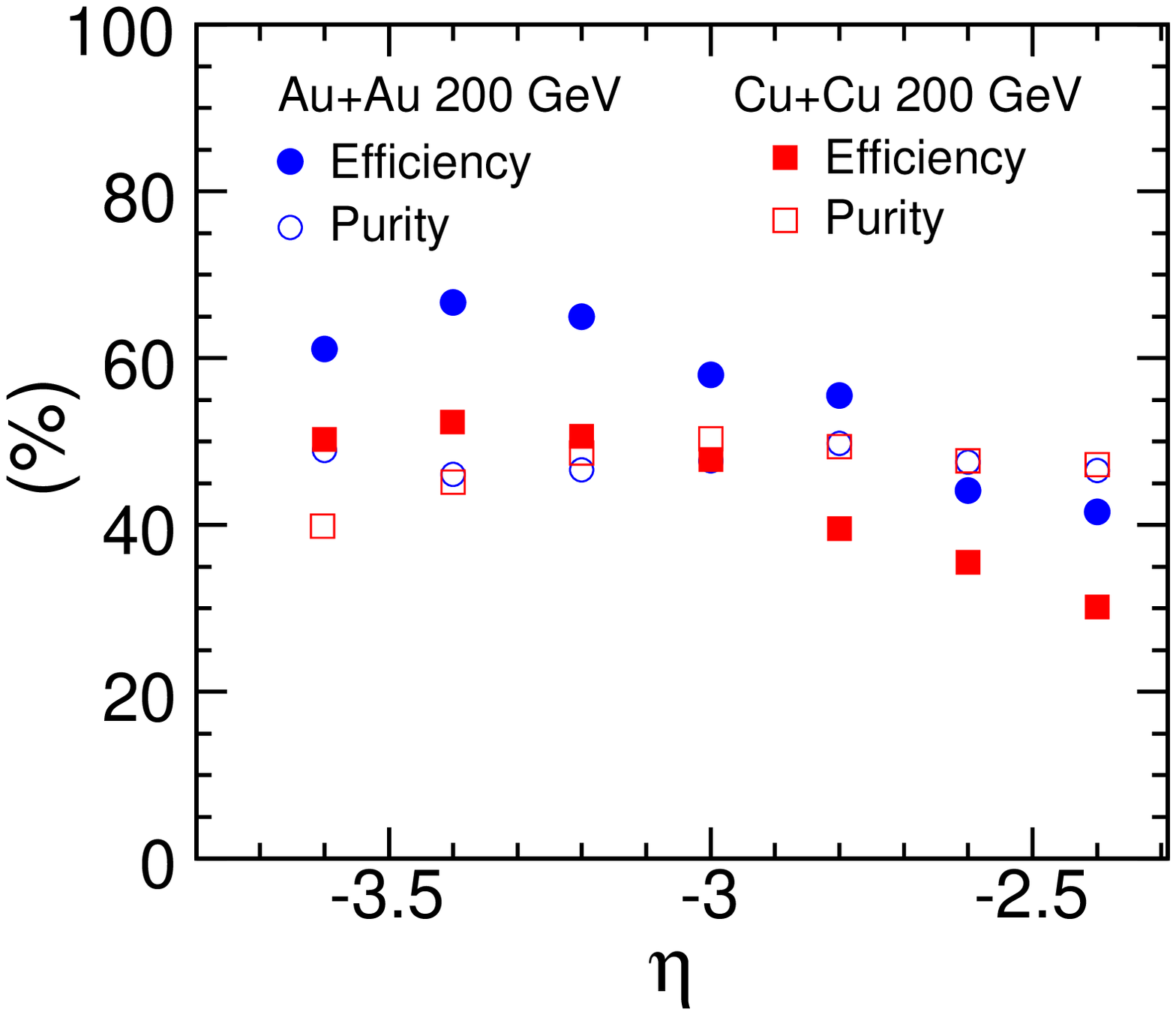}
\includegraphics[height=10cm,width=12cm]{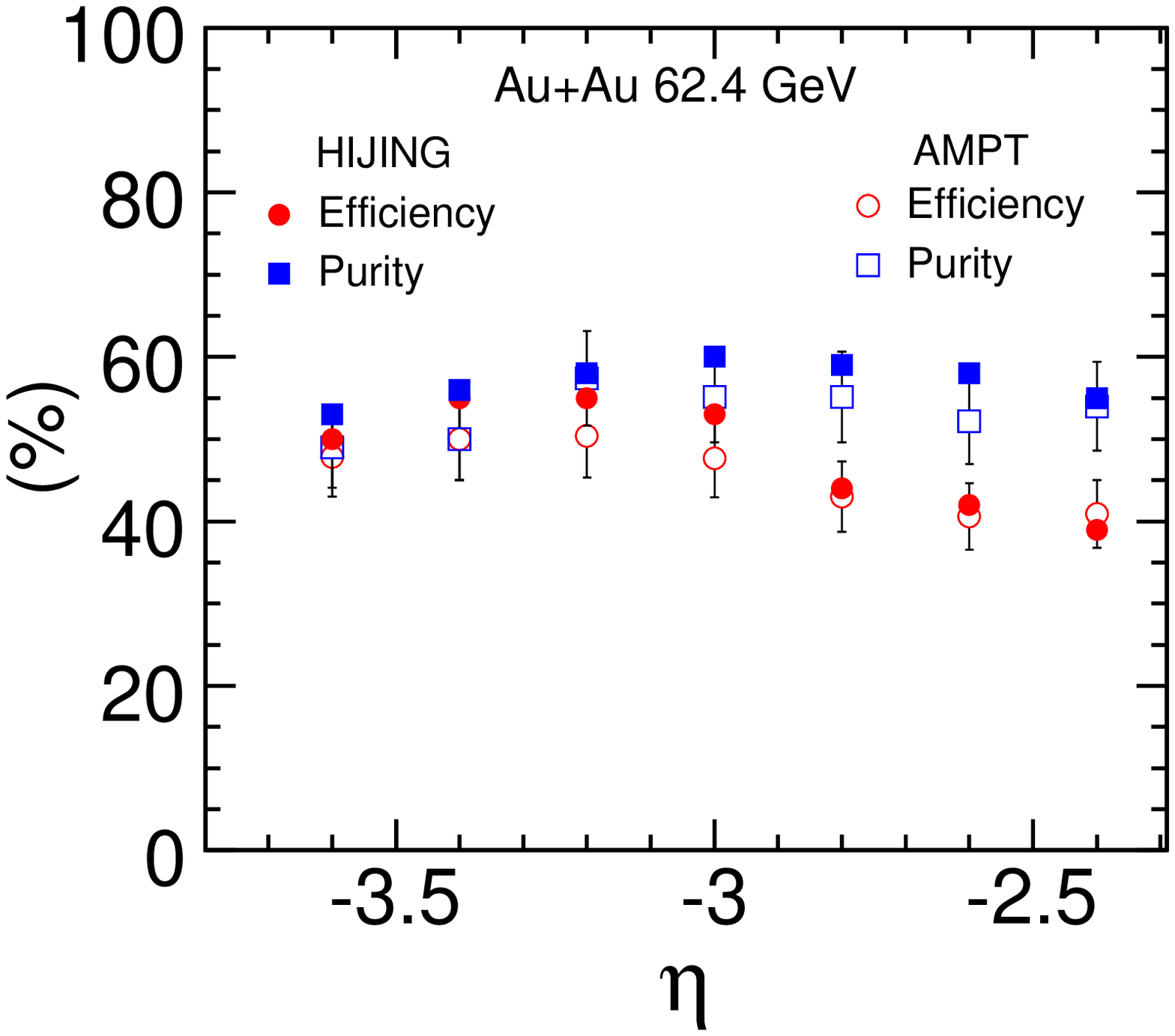}
\caption{ (color online) Top panel: Photon reconstruction efficiency ({\it $\epsilon_{\gamma}$}) (solid symbols) and purity of photon
sample ({\it $f_{\mathrm{p}}$}) (open symbols) for PMD as a function of pseudorapidity ($\eta$) for
minimum bias Au+Au and Cu+Cu at 
$\sqrt{s_{\mathrm {NN}}}$ = 200 GeV.
Bottom panel: Comparison between estimated {\it $\epsilon_{\gamma}$} and {\it $f_{\mathrm{p}}$} 
for PMD as a function of $\eta$ for minimum bias Au+Au at $\sqrt{s_{\mathrm {NN}}}$ = 62.4 GeV
using HIJING and AMPT models. The error bars on the AMPT data are statistical and those for
HIJING are within the symbol size.}
\label{fig0}
\end{center}
\end{figure*}

To estimate the number of photons ($N_{\mathrm {\gamma}}$) from the detected 
$N_{\mathrm {\gamma{\mathrm -like}}}$ clusters, we use the photon reconstruction
efficiency ({\it $\epsilon_{\mathrm {\gamma}}$}) and purity ({\it $f_{\mathrm {p}}$}) of 
the $\mathrm {\gamma{\mathrm -like}}$ sample defined~\cite{starphoton,wa98_dndy} 
as  $\epsilon_\gamma  =  N^{\mathrm {\gamma,th}} _{\mathrm cls} / N_\gamma$
and $f_{\mathrm {p}}  =  N^{\mathrm {\gamma,th}} _{\mathrm cls} / N_{\mathrm {\gamma{\mathrm -like}}}$, respectively.
$N_{\mathrm cls}^{\mathrm {\gamma,th}}$ is the number of photon clusters above the 
photon-hadron discriminator threshold. The photon multiplicity in data is then obtained
as $N_\gamma = (f_{\mathrm {p}}/\epsilon_{\mathrm {\gamma}}) N_{\mathrm {\gamma{\mathrm -like}}}$
~\cite{starphoton,starpmd_nim,wa98_dndy}, where the ratio $f_{\mathrm {p}}/\epsilon_{\mathrm {\gamma}}$
is estimated from simulations as described below.
Both {\it $\epsilon_{\gamma}$} 
and {\it $f_{\mathrm{p}}$} are obtained from a detailed Monte Carlo simulation using 
{\sc hijing} version 1.382~\cite{hijing} 
with default parameter settings and the detector simulation package 
{\sc geant}~\cite{geant}, which incorporates 
the full STAR detector framework. 
In our previous work~\cite{pmdftpc} it has been shown that HIJING reproduces the 
$N_{\rm{ch}}/N_{\gamma}$ 
ratio in Au+Au collisions at $\sqrt{s_{\rm{NN}}} =$ 62.4 GeV.
For estimation of {\it $\epsilon_{\gamma}$} in simulations it may be important
to know the inclusive photon $p_{\mathrm T}$ distribution.
Due to lack of experimental measurement of the inclusive photon $p_{T}$ distribution
at forward rapidity it assumed that they are similar to those from HIJING model. In order to
investigate the possible differences, {\it $\epsilon_{\gamma}$} and {\it $f_{\mathrm{p}}$}
are also obtained from a detailed Monte Carlo simulation using AMPT model~\cite{ampt} with default
parameter settings. The AMPT model is a multiphase transport model which includes both
initial partonic and final hadronic interactions.
The differences between the {\it $\epsilon_{\gamma}$} and {\it $f_{\mathrm{p}}$}
values estimated using the two models are less than 5\%. This difference is attributed to systematic errors
on $N_\gamma$.
Both {\it $\epsilon_{\gamma}$} and {\it $f_{\mathrm{p}}$} 
can vary with pseudorapidity and centrality.
The {\it $\epsilon_{\gamma}$} and {\it $f_{\mathrm{p}}$} for minimum bias Au+Au 
and Cu+Cu at 200 GeV are shown in top panel of Fig.~\ref{fig0}. 
The photon reconstruction efficiency (which includes the detector acceptance corrections) varies from 30\% 
at $\eta = -2.3$ to 60\% at $\eta = -3.7$ for all collision centralities obtained from simulations for 
Au+Au and Cu+Cu collisions at 62.4~\cite{starphoton,pmdftpc} and 200 GeV. The purity of the photon 
sample 
is nearly constant as a function of $\eta$ and varies between 40\% and 60\% for Au+Au 
and Cu+Cu collisions at 62.4~\cite{starphoton,pmdftpc} and 200 GeV. 
Both {\it $\epsilon_{\gamma}$} and {\it $f_{\mathrm{p}}$} show slight variation with system-size.
The $\eta$ dependence of the {\it $\epsilon_{\gamma}$} reflects mainly the varying 
detector acceptance 
between $\eta = -2.0$ and $\eta = -3.0$. There is also a small effect on the 
$\eta$ on {\it $\epsilon_{\gamma}$} due to varying particle 
density as a function of $\eta$. This is effect is already 
reflected in the comparison of {\it $\epsilon_{\gamma}$} values between 
Au+Au and Cu+Cu. The {\it $f_{\mathrm{p}}$} 
values by definition
are not affected by detector acceptance. The bottom panel of Fig.~\ref{fig0} 
shows a typical comparison of estimated {\it $\epsilon_{\gamma}$} and {\it $f_{\mathrm{p}}$}
using HIJING and AMPT models for Au+Au minimum bias collisions at 62.4 GeV. The differences
are within 5\% level. The systematic errors
are discussed below. The lower limit of photon $p_{\mathrm T}$ acceptance in the PMD is estimated
from detector simulations to be 20 MeV/$c$.

The systematic errors for photon multiplicity ($N_{\mathrm{\gamma}}$) 
are  due to~\cite{starphoton,pmdftpc} 
(a) uncertainty in estimates of  {\it $\epsilon_{\mathrm {\gamma}}$} and 
{\it $f_{\mathrm {p}}$}  values 
    arising from splitting of clusters, the choice of 
    photon-hadron discriminator threshold and choice of different event generators
    for their estimation and  
(b) uncertainty in $N_{\mathrm{\gamma}}$  arising from the 
non-uniformity of the detector response 
(primarily due 
to cell-to-cell gain variation). 
The error in $N_{\mathrm{\gamma}}$ due to (a) is
estimated from Monte Carlo simulations to be $\le$ 16\%
for all systems and beam energies studied. It is fairly
independent of collision centrality.
The error on $N_{\mathrm{\gamma}}$ due to (b) is
estimated using average gains for normalization and by studying the 
azimuthal dependence of the photon density, in an $\eta$ window in the detector,
and is found to be $\sim$ 10\%.
The total systematic error in $N_{\mathrm{\gamma}}$ is $\sim$19\% 
for both central and peripheral collisions and is similar for
Au+Au and Cu+Cu at 62.4 and 200 GeV.
The statistical errors are small and within the symbol size for the results
shown in the figures.


\begin{figure*}
\begin{center}
\includegraphics[scale=0.7]{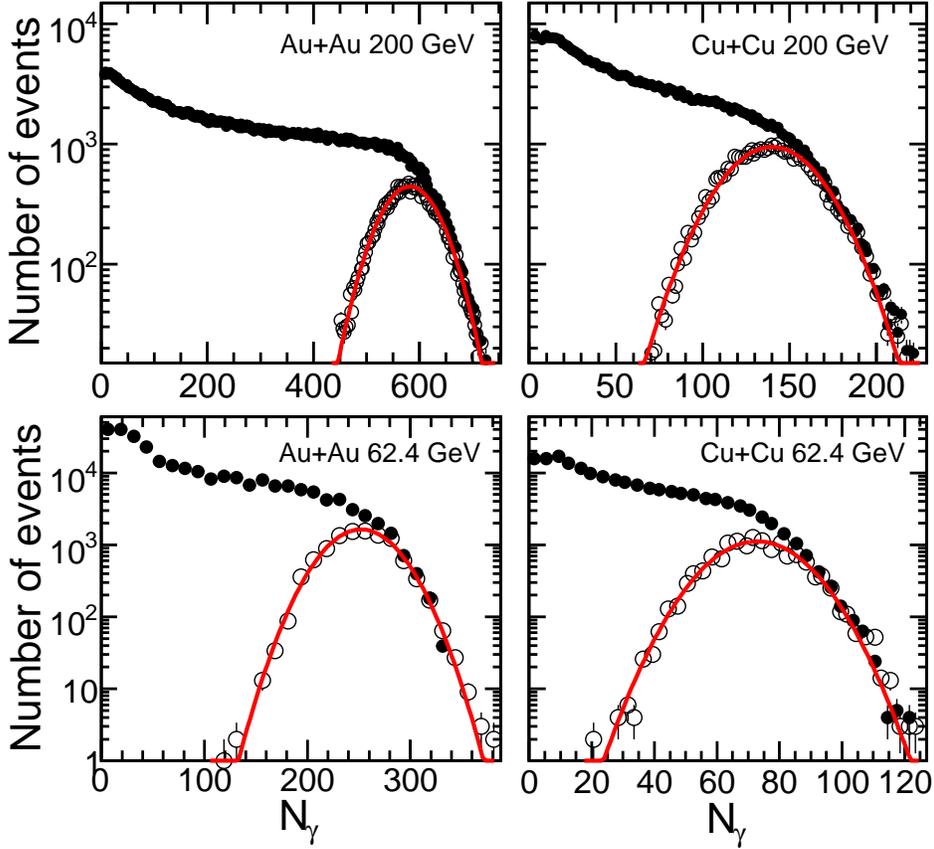}
\caption{ (color online) Event-by-event photon multiplicity distributions (solid circles) 
for Au+Au and Cu+Cu at 
$\sqrt{s_{\mathrm {NN}}}$ = 62.4 and 200 GeV. The distributions for top 0--5\% central 
Au+Au collisions and top 0--10\% central Cu+Cu collisions are also shown (open circles). 
The photon multiplicity distributions for central collisions are observed to be Gaussian (solid line). 
Only statistical errors are shown.}
\label{fig1}
\end{center}
\end{figure*}

\section{Multiplicity Distributions}

Figure~\ref{fig1} shows the photon multiplicity distributions 
for minimum bias Au+Au and Cu+Cu collisions at 62.4 and 200 GeV.
The distributions for both energies and colliding ion species 
show a characteristic shape with a rise at small multiplicity owing to peripheral
events. This indicates the probability of occurrence of peripheral collisions
is higher. This rise is followed by a near plateau region with increasing photon multiplicity.
This region is more prominent for Au+Au than Cu+Cu. 
It corresponds to mid-central collisions. There is a fall-off region in
the distributions for the most central collisions. The shape of the distribution
in the fall-off region is governed by intrinsic fluctuations in the measured
quantity and on the limited acceptance. Also shown in Fig.~\ref{fig1} are 
event-by-event photon multiplicity distributions for central Au+Au (0--5\%) 
and Cu+Cu (0--10\%) at $\sqrt{s_{\rm{NN}}}$ = 62.4 and 200 GeV. 
The solid lines are Gaussian fits to the data.
The fit parameters are given in Table~\ref{table1}. 

\begin{table}
\caption{\label{table1}
Gaussian fit parameters for photon multiplicity distributions for $-3.7 < \eta < -2.3$
for central Au+Au (0--5\%) and Cu+Cu (0--10\%)  at $\sqrt{s_{\mathrm {NN}}}$ = 62.4 and 200 GeV.}
\vspace{.5cm}
\begin{center}
\begin{tabular}{|c|c|c|c|c|}
\hline
Collision Type & $\langle N_{\mathrm {part}} \rangle$ & $\langle N_{\gamma} \rangle$ & $\sigma_{\gamma}$ \\
\hline
Au+Au 62.4 GeV &  347.3    &  252   & 30   \\
\hline
Au+Au 200 GeV  &  352.4    &  582   & 52    \\
\hline
Cu+Cu 62.4 GeV &  96.4     &  73    &  13    \\
\hline
Cu+Cu 200 GeV  &  99.0     &  140   &  26    \\
\hline
\end{tabular}
\end{center}
\end{table}

\begin{figure*}
\begin{center}
\includegraphics[scale=0.7]{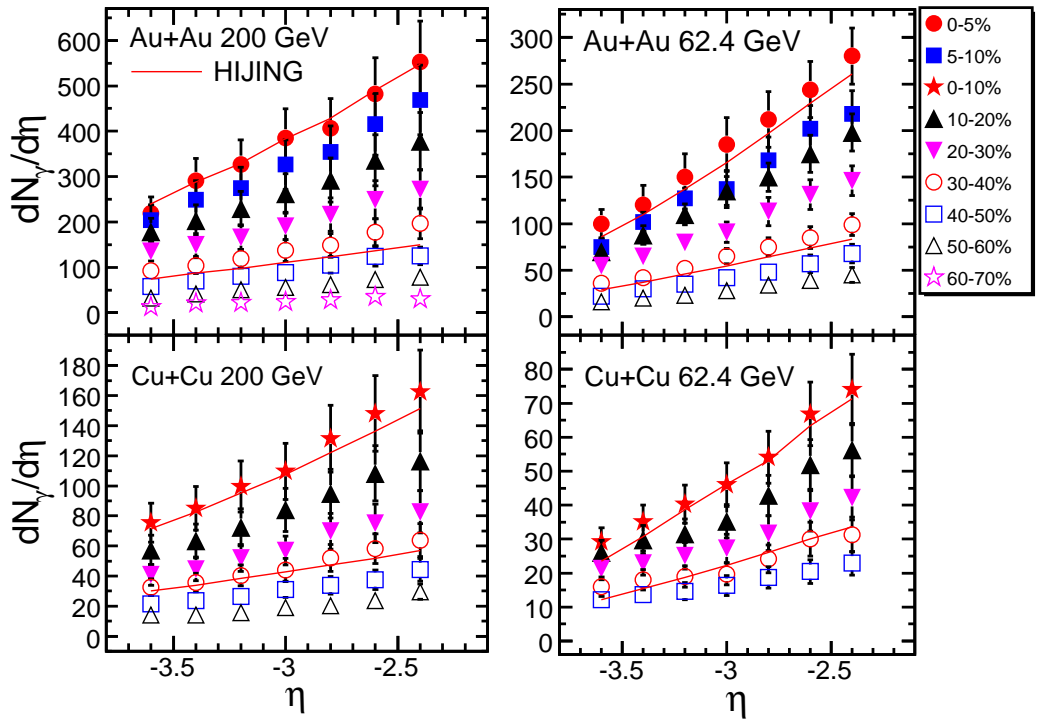}
\caption{ (color online) Photon pseudorapidity distributions for Au+Au and 
Cu+Cu at $\sqrt{s_{\mathrm {NN}}}$ = 62.4 and 200 GeV. 
The results for several centrality classes are shown. 
The solid curves are results of HIJING simulations for
central (0--5\% for Au+Au and 0--10\% for Cu+Cu) 
and 30--40\% mid-central collisions. 
The errors shown are systematic, statistical errors are negligible 
in comparison.}
\label{fig2}
\end{center}
\end{figure*}

Figure~\ref{fig2} shows the pseudorapidity distributions of photons measured in the PMD 
for various collision centralities in Au+Au and Cu+Cu at $\sqrt{s_{\mathrm {NN}}}$ = 62.4 and 
200 GeV. As expected, 
the photon yield increases with decreasing $|\eta|$. 
The photon multiplicity is found to increase 
from peripheral to central collisions. 
Comparisons to 
HIJING calculations for central (0--5\% for Au+Au and 0--10\% for Cu+Cu) and 30--40\% 
mid-central collisions are also shown in the figure (solid curves). 
The  HIJING results are in reasonable agreement with the data for both beam 
energies and colliding ion species. Similar conclusions are drawn for other centrality classes 
as well.


\section{Scaling of Photon Production}

\subsection{Scaling with $\langle N_{\mathrm{part}} \rangle$}
\begin{figure}
\begin{center}
\includegraphics[scale=0.6]{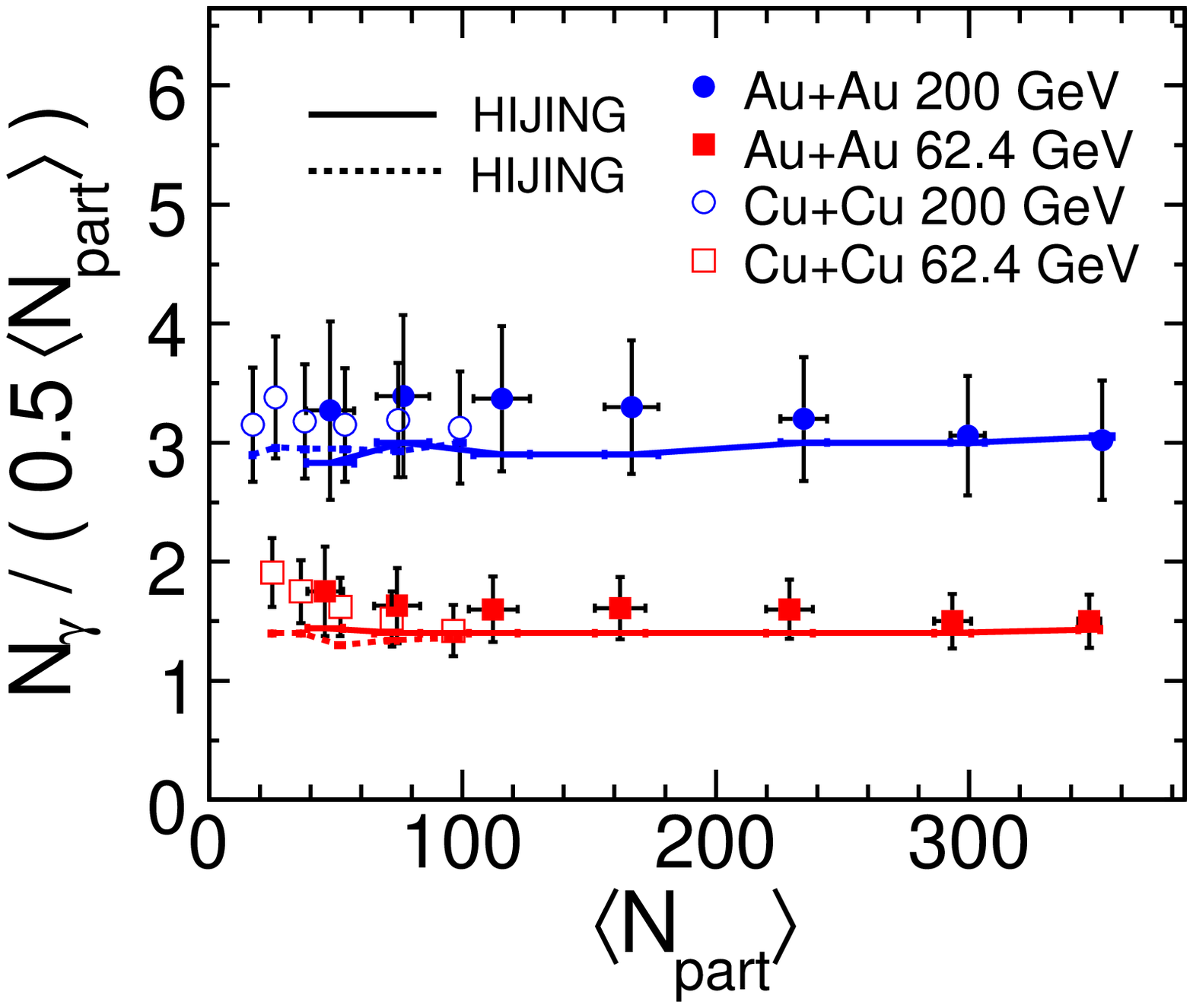}
\includegraphics[scale=0.6]{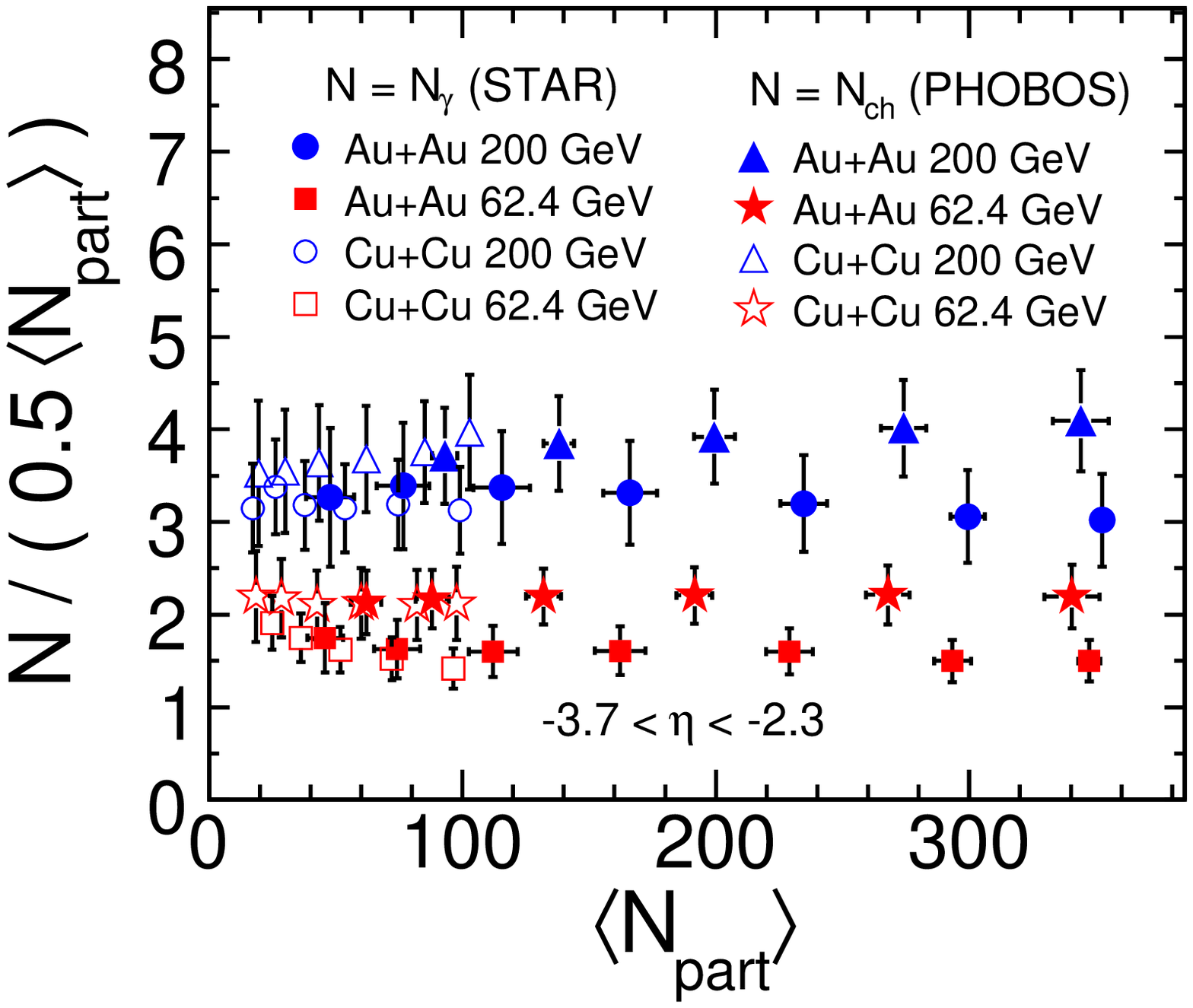}
\caption{ (color online) Top panel: The number of photons 
divided by $\langle N_{\mathrm{part}} \rangle/2$ as a function of average number 
of participating nucleons for Au+Au and Cu+Cu at  $\sqrt{s_{\mathrm {NN}}}$ = 62.4 
and 200 GeV for $-3.7 < \eta < -2.3$ . 
Errors shown are systematic only and include those for $\langle N_{\mathrm{part}} \rangle$.
Results from HIJING are shown as lines (solid for Au+Au 
and dashed for Cu+Cu).
Bottom panel: Same as above, for both photons and charged particles from PHOBOS~\cite{phobosscaling}.}
\label{fig3}
\end{center}
\end{figure}
The scaling of particle multiplicity with 
$\langle N_{\mathrm{part}} \rangle$
indicates the dominance of soft processes in particle production at RHIC, whereas scaling with average
number of binary collisions ($\langle N_{\mathrm{bin}} \rangle$) indicates the onset of hard processes (pQCD jets).
The PHENIX experiment first showed that at mid-rapidity, the charged particle production
scales 
with a combination of $\langle N_{\mathrm{part}} \rangle$ and $\langle N_{\mathrm{bin}} \rangle$~\cite{phenixscaling}, indicating 
significant
contribution of hard processes in particle production. The PHOBOS experiment showed that such 
scaling 
has a pseudorapidity dependence~\cite{phobosscaling}. At mid-rapidity ($|\eta|$ $<$ 1) 
particle production scales with a combination of $\langle N_{\mathrm{part}} \rangle$ and $\langle N_{\mathrm{bin}} \rangle$;
for the range 3 $<$ $|\eta|$ $<$ 3.4 it scales with  $\langle N_{\mathrm{part}} \rangle$; and for 
the region 5 $<$ $|\eta|$ $<$ 5.4, 
the particle production per average number of participating nucleon pair decreases with increasing $\langle N_{\mathrm{part}} \rangle$.

Figure~\ref{fig3} (top panel) shows the variation of photon multiplicity per average number of participating nucleon pair 
with $\langle N_{\mathrm{part}} \rangle$ for Au+Au and Cu+Cu at 62.4 and 200 GeV 
within the range $-3.7 < \eta < -2.3$. We observe that within the systematic errors, the 
photon multiplicity scales with $\langle N_{\mathrm{part}} \rangle$ at forward rapidities. This indicates
that the photon production at forward rapidities is due to soft processes.
For collisions with similar $\langle N_{\mathrm{part}} \rangle$, the photon multiplicity is similar for 
Au+Au and Cu+Cu at a given beam energy.  Also shown in the figure are results
from HIJING (solid lines for Au+Au and dashed lines for Cu+Cu).
Considering the systematic errors shown, the HIJING results compare well 
with the data for most of the collision centralities studied.

Figure~\ref{fig3} (bottom panel) shows the comparison of photon multiplicity per average number of participating
nucleon pair vs. $\langle N_{\mathrm{part}} \rangle$ and the corresponding data for charged particles from the
PHOBOS experiment 
for the range $-3.7 < \eta < -2.3$. Like photon production, the charged particle 
multiplicity at forward rapidities is found to scale with $\langle N_{\mathrm{part}} \rangle$. For similar 
$\langle N_{\mathrm{part}} \rangle$, 
the charged particle production in the region $-3.7 < \eta < -2.3$ is also found to be similar for
Au+Au and Cu+Cu at a given beam energy. The photon production per 
average number of participating nucleon pair is 
slightly lower compared to that for charged particles. 
A constant straight line combined fit to the charged particle 
results for Au+Au and Cu+Cu in Fig.~\ref{fig3} at  $\sqrt{s_{\rm{NN}}}$ = 200 GeV gives 
3.8 $\pm$ 0.2,
while that for photons yields 
3.2 $\pm$ 0.1. 
For $\sqrt{s_{\rm{NN}}}$ = 62.4 GeV the values are 
2.2 $\pm$ 0.1 and 
1.6 $\pm$ 0.05 for charged particles and photons, respectively. 
The difference may be due to the contribution of protons to
charged particles at forward rapidity. The measurements ($-3.7 < \eta < -2.3$)
are carried out close to the fragmentation region, where protons play 
an increasingly larger role~\cite{pmdftpc,brahms_proton}.
The ratio of the number of charged particles to photons in the range $-3.7 < \eta < -2.3$ is found 
to be 1.4 $\pm$ 0.1 and 1.2 $\pm$ 0.1 for $\sqrt{s_{\mathrm NN}}$ = 62.4 GeV and 200 GeV, respectively.

\begin{figure}
\begin{center}
\includegraphics[scale=0.6]{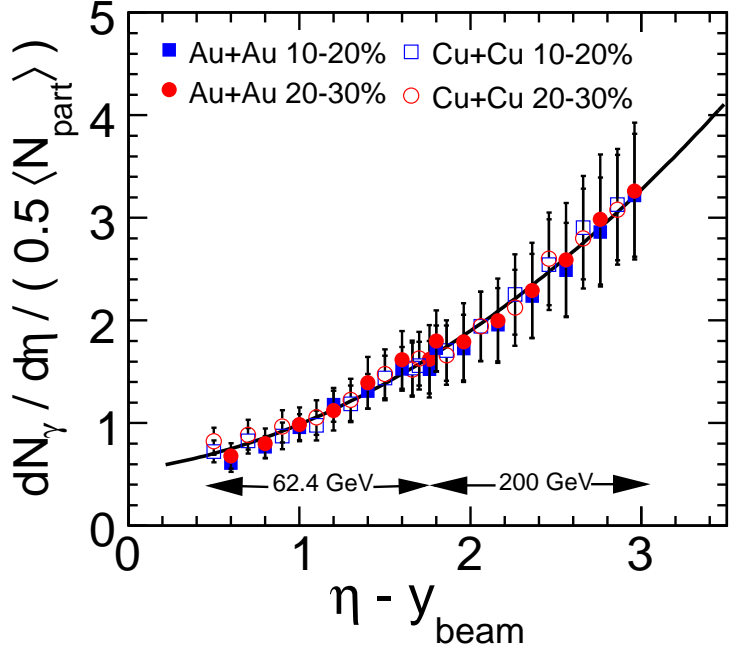}
\includegraphics[scale=0.6]{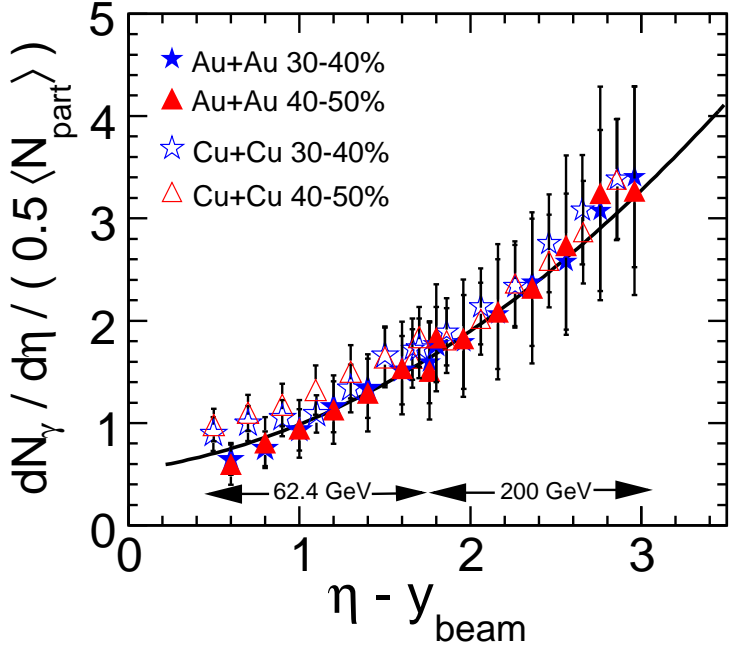}
\caption{ (color online) Photon pseudorapidity distributions normalized by the average 
number of participating nucleon pairs
for different collision centralities are plotted as a function of pseudorapidity shifted by the 
beam rapidity ($-5.36$ for 200 GeV and $-4.19$ for 62.4 GeV) for Au+Au and Cu+Cu collisions at 
$\sqrt{s_{\rm{NN}}} =$  62.4 and 200 GeV. 
Errors are systematic only, statistical errors are negligible in comparison. For clarity of presentation, 
results for only four 
centralities are 
shown. The Cu+Cu data are shifted by 0.1 unit 
in $\eta-y_{\rm {beam}}$. 
The solid line is a second order polynomial fit to the data (see text for details).}
\label{fig4}
\end{center}
\end{figure}

\subsection{Longitudinal Scaling}

Previously it was reported that both charged particle~\cite{phobosscaling,brahms} 
and photon pseudorapidity density~\cite{starphoton,pmdftpc},
normalized by the average number of participating nucleon pairs as a function of $\eta-y_{\rm {beam}}$,
where $y_{\mathrm beam}$ is the beam rapidity, is independent of beam energy. Further, it was observed that
such longitudinal scaling was centrality dependent for charged particles, but was centrality 
independent 
for photons~\cite{starphoton,pmdftpc}. Figure~\ref{fig4} shows the photon pseudorapidity 
density normalized by the average number of 
participating nucleon pairs as a function of 
$\eta-y_{\rm {beam}}$, for selected
centralities (for the sake of clarity) for Au+Au and Cu+Cu at
$\sqrt{s_{\rm{NN}}}$ = 62.4 and 200 GeV. The $y_{\mathrm {beam}}$ values for 62.4 and 200 GeV are 
$-4.19$ and $-5.36$, respectively. The Cu+Cu results are shifted by 0.1 units in $\eta$ for sake of 
clarity. 
The solid line is a second order polynomial of the form 
0.54 + 0.22($\eta-y_{\rm {beam}}$) + 0.23($\eta-y_{\rm {beam}})^{2}$, fitted to all the 
data of Fig.~\ref{fig4}. A fit to the ratio of  data to this function for the 
results in the
upper panel yields a value of 0.96 $\pm$ 0.01 and those on the lower panel yields 
1.03 $\pm$ 0.01.
The results demonstrate that the longitudinal scaling for produced photons is 
independent
of colliding ion species. In addition we re-confirm that such scaling for photons is
independent of beam energy and collision centrality as reported earlier~\cite{starphoton,pmdftpc}.

\section{Summary}

Photon multiplicity distributions are measured at forward rapidity ($-3.7 < \eta < -2.3$)
for Au+Au and Cu+Cu collisions at $\sqrt{s_{\rm{NN}}}$ = 62.4
and 200 GeV using the photon multiplicity detector in the STAR experiment at RHIC. 
As expected, the photon yield increases with decreasing 
$|\eta|$
(towards mid-rapidity), and is larger for collisions 
at higher  energies. The photon multiplicity per participating nucleon pair is observed to be 
independent of collision centrality indicating that photon production is dominated by soft 
processes. A similar observation is made for charged particles, although their production 
is slightly higher.
This slightly higher production of charged particles than photons could be 
due to the contribution of baryons 
to the charged particles, which may come from baryon transport and contribution from beam protons.
On the other hand, photons are mainly from the decay of mesons ($\pi^{0}$).
For collisions with similar 
average number of participating nucleons, the photon yields are similar for Au+Au and Cu+Cu 
for a given colliding beam energy. 
The photon production per unit rapidity per 
average number of participating nucleon pair vs. $\eta-y_{\mathrm beam}$ 
shows longitudinal scaling which is independent of beam energy, collision centrality 
and colliding ion species.

We thank the RHIC Operations Group and RCF at BNL, the NERSC Center at LBNL 
and the Open Science Grid consortium for providing resources and support. 
This work was supported in part by the Offices of NP and HEP within the U.S. 
DOE Office of Science, the U.S. NSF, the Sloan Foundation, the DFG cluster of 
excellence `Origin and Structure of the Universe', CNRS/IN2P3, STFC and EPSRC 
of the United Kingdom, FAPESP CNPq of Brazil, Ministry of Ed. and Sci. of the 
Russian Federation, NNSFC, CAS, MoST, and MoE of China, GA and MSMT of the 
Czech Republic, FOM and NWO of the Netherlands, DAE, DST, and CSIR of India, 
Polish Ministry of Sci. and Higher Ed., Korea Research Foundation, Ministry 
of Sci., Ed. and Sports of the Rep. Of Croatia, Russian Ministry of Sci. 
and Tech, and RosAtom of Russia.


\end{document}